# Dispersion Equation for Smith-Purcell FEL

D.N. Klochkov1 , M.A. Kutlan 2*
1 Prokhorov General Physics Institute, Moscow, Russia
2 Institute for Particle & Nuclear Physics, Budapest, Hungary

* kutlanma@gmail.com

**Abstract:** For the grating, which has depth of grooves as a small parameter, the dispersion equation of the Smith-Purcell instability was obtained. It was found that the condition of the Thompson or the Raman regimes of excitation does not depend on beam current but depends on the height of the beam above grating surface. The growth rate of instability in both cases is proportional to the square root of the electron beam current.

1. **Overview**

Several theories have been proposed to describe the operation of a Smith-Purcell FEL. In particular, Schaechter and Ron [1] proposed a theory based on the interaction of an electron-beam with a wave traveling along the grating. The interaction is found to amplify the waves that are incident on the grating and reflected by it, with a gain that depends on the reflection matrix of the grating. They have found that the gain is proportional to the cube root of the electron-beam current, which with the behavior of Cherenkov free-electron lasers and other slow-wave devices. More recently, Kim and Song [2] have proposed a theory in which they assume that the electrons interact with a wave that travels along the surface of the grating. Assuming that at least one Fourier component of the traveling wave is radiative, they have found that the gain is proportional to the square root of the electron-beam current $I_b^{1/2}$.

Previous theories of the SP-FEL have assumed that the electron beam interacts with a wave the frequency of which is the same as that of the SP radiation. It has recently been proposed that the beam interacts with an evanescent mode of the grating that lies at a wavelength longer than the SP radiation and radiates only when it reaches the end of the grating [3-5]. When the group velocity v_g of this mode is positive, the interaction corresponds to a convective instability and feedback must be provided by an external resonator (or reflections from the ends of the grating). When the group velocity of the evanescent mode is negative, the interaction corresponds to an absolute instability and the SP-FEL oscillates without external feedback if the current is above a threshold value called the start current.

A rectangular grating is considered in the papers [3-5], assuming that the entire space above the grating is filled by a uniform electron beam. The authors of [3-5] have established the dispersion relation and found the dispersion law **k.** It turns out that the dispersion equation allows only evanescent solutions and the operating point of a Smith-Purcell free electron laser is fixed by the intersection of the dispersion curve with the beam line. The corresponding small signal gain follows the $I_b^{1/3}$ law.

There are numerous publications devoted to FELs on undulators and strophotrons [7-49 and references therein]. After the advent of Free Electron Lasers (FEL), it was suggested to use the Smith-Purcell effect to create a new type of FEL – the Smith-Purcell Free Electron Laser (SP FEL).

In the present article we consider a simple model of grating and obtain the dispersion equation for the SP-system. Assuming that the depth of the grating groove h is a small parameter permits us to solve analytically the dispersion equation. We consider a general model for a metal grating and obtain the dispersion equation and its solutions for an arbitrary grating profile.

## 2. Dispersion equation: particular case

The solutions obtained in the [6] define the amplitudes of the field but do not determine its frequency and the growth rates of SP instability. In order to get the frequency we have to consider the boundary condition on the grating surface.

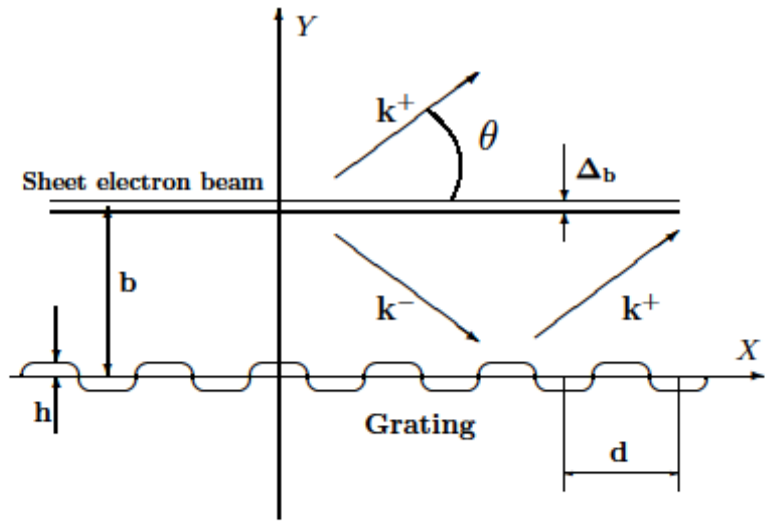

Figure 1. Schematic of a SP-FEL using a sheet electron beam. The sheet electron beam is in the plane $y = b$. $\theta$ is the angle of observation, $\Delta_b$ is the thickness of the beam, $d$ and $h$ are the period and the amplitude of the grating, respectively.

As a first step, we consider the simplest model of the grating, when the equation of metal surface is defined as

$$y(x) = h\sin(\chi x) = h\sin\left(\frac{2\pi}{d}x\right). \qquad (1)$$

The vector tangential to the surface is $\mathbf{r}_0 = (1, h\chi\cos(\chi x))$. The boundary condition $(\mathbf{r}_0 \mathbf{E}(x, y(x))) = 0$ takes the form

$$E_x(x, y(\mathrm{x})) + h\chi\cos(\chi x)E_y(x, y(\mathrm{x})) = 0. \qquad (2)$$

Here $E_x$ and $E_y$ are given by (11) of [6]. Substituting the solutions of Maxwell equations given by (10) and (23) of [6] in boundary condition (2), multiplying by $exp(-ip\chi x)$, where $p = 0; \pm1; \pm2; \pm3,...$, and integrating then over coordinate x within the interval $(0, d)$ we obtain an infinite system of algebraic equations for the partial amplitudes $P_n$ (see Appendix A):

$$\sum_n \left(q_{ny} + (n-p)\chi\frac{q_{nx}}{q_{ny}}\right)\left(J_{n-p}(-q_{ny}h) + K(n)\left[J_{n-p}(-q_{ny}h) - J_{n-p}(q_{ny}h)e^{2iq_{ny}b}\right]\right)P_n = 0 \qquad (3)$$

Here $J_\alpha(x)$ are a Bessel functions.

### a) 3-waves approximation

In order to cut the infinite system we assume that the amplitude of the grating h is vary small. Since the Bessel function of the small variable x drops as $J_\alpha(x) \approx (x/2)^\alpha/\alpha!$ with $\alpha$ increasing, the assumption that $h$ is a small parameter permits us to consider 3-waves approximation. We suppose that there are only three non-zero partial amplitudes, namely, $P_{n-1}$, $P_n$ and $P_{n+1}$. Other amplitudes are assumed to be zero. Then we get three algebraic equations

$$\mathbf{MV} = 0 \qquad (4)$$

where $\mathbf{V}$ is a column vector with components $P_{n-1}$, $P_n$ and $P_{n+1}$, $\mathbf{M}$ is a $3\times3$ matrix expressed as

$$\begin{Vmatrix} q_{n-1y}\alpha_{n-1}, \beta_n^+, \ldots\ldots, \gamma_{n+1}^+ \\ \beta_{n-1}^-, \ldots, q_{ny}\alpha_n, \ldots, \beta_{n+1}^+ \\ \gamma_{n-1}^-, \ldots\ldots, \beta_n^-, \ldots, q_{n+1y}\alpha_{n+1} \end{Vmatrix}. \tag{5}$$

Here the coefficients are

$$\alpha_n = 1 + K(n)\left(1 - e^{2iq_{ny}b}\right),$$
$$\beta_n^\pm = \mp\frac{h}{2}\left(q_{ny}^2 \mp \chi q_{nx}\right)\left[1 + K(n)\left(1 + e^{2iq_{ny}b}\right)\right], \tag{6}$$
$$\gamma_n^\pm = \frac{h^2}{8}q_{ny}\left(q_{ny}^2 \pm 2\chi q_{nx}\right)\alpha_n.$$

The nontrivial solution of the finite system of equations (4) takes place if its determinant equals zero. This dispersion equation describes both the spectrum of frequencies and the growth rates of the stimulated Smith-Purcell radiation. Neglecting the small terms proportional to $h^p$ with $p \geq 4$, we obtain the following simple dispersion equation:

$$D(\omega,\mathbf{k}) = q_{n-1y}q_{ny}q_{n+1y}\alpha_{n-1}\alpha_n\alpha_{n+1} - q_{n-1y}\alpha_{n-1}\beta_n^-\beta_{n+1}^+ - \beta_{n-1}^-\beta_n^+ q_{n+1y}\alpha_{n+1} = 0 \tag{7}$$

The question is which the zero-order approximation for dispersion equation is. In the absence of the beam, when $\omega_b = 0$, we find from Eq. (23) of [6] that $P_n^- = 0$ and $P_n^+ = P_n$, which means that $\mathbf{E}_n^- = 0$ and $\mathbf{E}_n^+ = \mathbf{E}_n$. Substituting this amplitudes in the boundary condition we find that $\mathbf{E}_n = 0$. Since all amplitudes are zero, the determinant of the system is not equal to zero. Indeed, substituting $\omega_b = 0$ in equation (7) we get

$$D(\omega,\mathbf{k},\omega_b = 0) = q_{n-1y}q_{ny}q_{n+1y}$$
$$+ \frac{h^2}{4}\left\{q_{n-1y}\left(q_{ny}^2 - \chi q_{nx}\right)\left(q_{n+1y}^2 + \chi q_{n+1x}\right) + q_{n+1y}\left(q_{n-1y}^2 - \chi q_{n-1x}\right)\left(q_{ny}^2 + \chi q_{nx}\right)\right\} \neq 0. \tag{8}$$

It has a simple physical explanation. In the absence of the beam, the electromagnetic field in the system is absent too, because the grating does not emit the waves.

Therefore we can not use the condition of $\omega_b = 0$ as a zero-order approximation to solve the dispersion equation. Instead we use as zero-order approximation condition of $h = 0$, which gives

$$D(\omega,\mathbf{k},h=0) = q_{n-1y}q_{ny}q_{n+1y}\alpha_{n-1y}\alpha_{ny}\alpha_{n+1y} = 0. \tag{9}$$

In the general case for an arbitrary angle $\theta$ the product of $q_{n-1y}q_{ny}q_{n+1y}$ does not equal zero, therefore we can write

$$D_0(\omega,\mathbf{k}) = \alpha_{n-1y}\alpha_{ny}\alpha_{n+1y} = 0 \tag{10}$$

Without loss of generality we assume that the $n$-mode of electromagnetic spectrum excites in the system. It means that $\alpha_n$, which gives the following equation

$$(\omega - (k_x + n\chi)u)^2 + \frac{i}{2}\Delta_b \omega_b^2 \gamma_0^{-3} q_{ny}\left(1 - e^{2iq_{ny}b}\right) = 0. \tag{11}$$

We seek the solution of the equation in the form

$$\omega = \omega_n \pm \Omega_{bn}, \tag{12}$$

where $\omega_n$ satisfies the equation (11) in the limit $\omega_b \to 0$:

$$\omega_n = (k_x + n\chi)u, \tag{13}$$

which is the dispersion relation for longitudinal beam waves.

Taking into account the dispersion relation for electromagnetic wave in vacuum

$$k_x = \frac{\omega_n}{c}\cos\theta, \tag{14}$$

we find the frequency of the $n$-mode of the spectrum for the limit $\omega_b \to 0$ as the intersection point in the $k_x, \omega$ diagram shown in the Fig.2.

$$\omega_n = \frac{n\chi u}{1 - \beta\cos\theta}, \tag{15}$$

where $\beta = u/c$ is the dimensionless velocity of electron, $\theta$ is the angle between $\mathbf{k}^+$ vector and positive direction of $x$-axis, i.e. the angle of observation. Please do not confuse the $n$-mode in the spectrum of radiation with the $n$-mode in $q_x$-spectrum of surface wave $E_n$. With the help of Eq. (15), from Eq. (8) of [6] we find that

$$q_{ny} = i\frac{\omega_n}{u\gamma_0}, \tag{16}$$

So the $n$-mode of the surface wave $E_n$ does not radiate, because it damps in the positive direction of the $y$-axis.

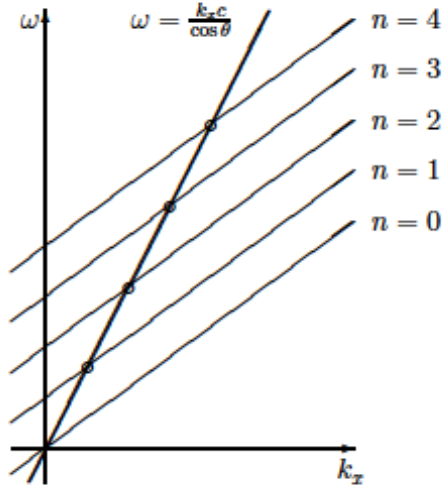

Fig.2. The dispersion relations for the electromagnetic wave $\omega = k_x c / \cos\theta$ and the beam waves with $\omega_n = (k_x + n\chi)u$, are shown in $k_x, \omega$ plane. The intersection points determine the spectrum of frequency of Eq. (15).}.

Considering the presence of beam as perturbation and substituting the value of $q_{ny}$ from Eq. (16) in Eq. (11) we can derive the beam frequency $\Omega_{bn}$:

$$\Omega_{bn} = \omega_b \gamma_0^{-2} \sqrt{\frac{\Delta_b \omega_n}{2u}\left(1 - e^{-2\frac{\omega_n b}{u\gamma_0}}\right)}. \tag{17}$$

Here we give the physical interpretation of the solutions obtained. The condition h=0 means that there is a flat mirror in the plane $y$. The fluctuations of charge and current densities of the electron beam produce fluctuations of the electromagnetic field, the plane wave of which is reflected from the mirror. The resonant condition (15) is the crossing point of the dispersion lines for the beam wave and the reflected electromagnetic plane wave. The boundary condition yields a discrete spectrum, the modes of which contain two coupled frequencies: ``high'' frequency with $\omega = \omega_n + \Omega_{bn}$ and ``low'' frequency with $\omega = \omega_n - \Omega_{bn}$. The beam frequency $\Omega_{bn}$ depends on the height of the beam above the surface of $y = 0$ and has the following asymptotes:

$$\Omega_{bn} = \begin{cases} \omega_b \gamma_0^{-2} \sqrt{\dfrac{\Delta_b \omega_n}{2u}},\,\,\,\,\,\,b \to \infty \\ \omega_b \gamma_0^{-5/2} \dfrac{\omega_n}{2u} \sqrt{\Delta_b b},\,\,\,\,b \to 0. \end{cases} \quad (18)$$

Now we shall consider the presence of grooves on the mirror ($h \ne 0$) as a perturbation and formulate the solution of the dispersion equation near the synchronism point (12) as:

$$\omega = \omega_n \pm \Omega_{bn} + \delta\omega, \quad (19)$$

where $\delta\omega = \delta\omega'_n + i\delta\omega''_n$ is a complex number, and the imaginary part $\delta\omega''_n$ is the growth rate of the induced Smith-Purcell instability. Substituting solution (19) in Eq. (7) and neglecting the small terms under the realistic condition of $\{\omega_b, \Omega_{bn}, |\delta\omega|\} \Box \chi u$ we obtain the dispersion equation for small shift of frequency:

$$\delta\omega^2 \pm 2\Omega_{bn}\delta\omega = \dfrac{h^2}{4} \dfrac{\Delta_b \omega_b^2}{\gamma_0^5} \left(\dfrac{\omega_n}{u}\right)^3 e^{-2\frac{\omega_n b}{u\gamma_0}} \left[X_n(\theta,u) + iY_n(\theta,u)\right]. \quad (20)$$

For $\omega = \omega_n$ the wave number $q_{n+1y}$ is always an imaginary value. We rewrite $q_{n+1y} = ig_{n+1}$,

Where $g_{n+1} = \dfrac{\omega_n}{nu} \sqrt{(n+1-\beta\cos\theta)^2 - n^2\beta^2}$. The wave number $q_{n-1y}$ can be both real and imaginary for $\omega = \omega_n$. Under the condition

$$\dfrac{1-\beta\cos\theta}{1+\beta} < n < \dfrac{1-\beta\cos\theta}{1-\beta} \quad (21)$$

the wave number $q_{n-1y}$ is real and then the coefficients

$$X_n = \left(1 + \dfrac{\chi u \gamma_0^2}{\omega_n}\right)\left(\dfrac{\chi}{g_{n+1}}\left(1 + \dfrac{\chi u}{\omega_n}\right) - \dfrac{g_{n+1} u}{\omega_n}\right),$$
$$Y_n = \left(1 - \dfrac{\chi u \gamma_0^2}{\omega_n}\right)\left(\dfrac{q_{n-1} u}{\omega_n} - \left(1 - \dfrac{\chi u}{\omega_n}\right)\dfrac{\chi}{q_{n-1y}}\right). \quad (22)$$

The other case is real under the condition

$$n \notin \left(\dfrac{1-\beta\cos\theta}{1+\beta}, \dfrac{1-\beta\cos\theta}{1-\beta}\right), \quad (23)$$

when the wave number $q_{n-1y} = ig_{n-1}$ is imaginary and then the coefficients

$$X_n = \left(1 + \frac{\chi u \gamma_0^2}{\omega_n}\right)\left(\frac{\chi}{g_{n+1}}\left(1 + \frac{\chi u}{\omega_n}\right) - \frac{g_{n+1} u}{\omega_n}\right) - \left(1 - \frac{\chi u \gamma_0^2}{\omega_n}\right)\left(\frac{g_{n-1} u}{\omega_n} - \left(1 - \frac{\chi u}{\omega_n}\right)\frac{\chi}{g_{n-1}}\right), \quad (24)$$

$$Y_n = 0.$$

here $g_{n-1} = \frac{\omega_n}{nu}\sqrt{(n+1+\beta\cos\theta)^2 - n^2\beta^2}$.

The quadratic equation (20) has simple solutions. Here we consider two interesting cases. First, we shall consider the Thompson type of excitation in a single-particle approximation, when the electron beam radiates as a single particle or a bunch.

In this case the beam waves can be neglected and the condition of generation is $|\delta\omega| \gg \Omega_{bn}$. Under this condition we can omit the second term in the left-hand side of the dispersion equation and obtain the solution:

$$\delta\omega_n = \frac{h\omega_n}{2u}\sqrt{\frac{\Delta_b \omega_n}{u}} \frac{\omega_b}{\gamma_0^{5/2}} e^{-\frac{\omega_n b}{u\gamma_0}} (X_n + iY_n)^{1/2}. \quad (25)$$

The growth rate of instability of the n-mode (15) ($\delta\omega = \delta\omega_n' + i\delta\omega_n''$) in this case is

$$\delta\omega_n'' = \pm \frac{h\omega_n}{2u} \frac{\omega_b}{\gamma_0^{5/2}} \sqrt{\frac{\Delta_b \omega_n}{u}} e^{-\frac{\omega_n b}{u\gamma_0}} (X_n + iY_n)^{1/4} \sin\left|\frac{\psi}{2}\right|, \quad (26)$$

where angle $\psi$ is defined as $\psi = \arccos\left(X_n / \sqrt{X_n^2 + Y_n^2}\right)$.

For sufficiently large value of the relativistic factor $\gamma_0$, when the inequality $\gamma_0 \gg \omega_n b / u$ holds, the growth rate of instability is proportional to the square root of inverse $\delta\omega'' \propto \gamma_0^{-1/2}$. The condition of the Thompson regime excitation does not depend on the beam current; it has the following form:

$$\frac{h\omega_n}{2u} \frac{|X_n + iY_n|}{\sqrt{2\gamma_0 \left(e^{2\frac{\omega_n b}{u\gamma_0}} - 1\right)}} \gg 1. \quad (27)$$

This inequality holds for $b \to 0$ and for large values of the grooves depth $h$ of the grating.

The second case is the Raman (or collective) regime of generation, when the influence of beam waves is appreciable and $|\delta\omega| \ll \Omega_{bn}$. For the Raman regime we obtain

$$\delta\omega = \pm \frac{h^2 \omega_n^2}{4u^2} \frac{\omega_b}{\gamma_0^3} \sqrt{\frac{\Delta_b \omega_n}{2u}} \frac{e^{-2\frac{\omega_n b}{u\gamma_0}}}{\sqrt{1-e^{-2\frac{\omega_n b}{u\gamma_0}}}} (X_n + iY_n). \qquad (28)$$

The growth rate of instability is

$$\delta\omega_n'' = \pm \frac{h^2 \omega_n^2}{4u^2} \frac{\omega_b}{\gamma_0^3} \sqrt{\frac{\Delta_b \omega_n}{2u}} \frac{e^{-2\frac{\omega_n b}{u\gamma_0}}}{\sqrt{1-e^{-2\frac{\omega_n b}{u\gamma_0}}}} Y_n(\theta, u). \qquad (29)$$

If $Y_n(\theta,u)$, then the high-frequency branch with $\omega = \omega_n + \Omega_{bn}$ excites. In the opposite case the low-frequency waves with $\omega = \omega_n - \Omega_{bn}$ generate.

For sufficiently large value of the relativistic factor $\gamma_0$, when the inequality $\gamma_0 \gg \omega_n b/u$ holds, the growth rate of instability is proportional to inverse $\gamma_0$: $\delta\omega'' \sim \gamma_0^{-1}$. The condition for the Raman type of excitation is

$$\frac{h^2 \omega_n^2}{4u^2 \gamma_0} \frac{|Y_n|}{e^{2\frac{\omega_n b}{u\gamma_0}} - 1} \ll 1. \qquad (30)$$

It does not depend on Langmuir beam frequency (or beam current), but it depends on the beam height $b$ above the grating.

The mechanism of light radiation is as follows. The surface wave with mode number $n$ acting on the beam electrons perturbs the trajectories of the latter. Electrons, oscillating near the equilibrium point of $\mathbf{r} = \mathbf{r}_0 + \mathbf{u}t$, emit electromagnetic waves. As, the amplitude of oscillation is proportional to the magnitude of the surface wave at the beam height $b$, the value of the growth rate depends on the beam height above the grating surface.

**b). 5-waves approximation**

In the previous subsection we considered the 3-waves approximation. When $q_{n-1y} = 0$ the 3-waves approximation does not work. It can be when $n \approx (1-\beta\cos\theta)(1\pm\beta)$, for example, for $n=1$ and $\theta=0$. In this case we have to include additional surface modes and consider 5-waves approximation; we assume that there are only five non-zero amplitudes: $E_{n-2x}$, $E_{n-x}$, $E_{nx}$ and

$E_{n+x}$ and $E_{n+2x}$. Other amplitudes are assumed to be zero. Taking into account that $h$ is a small parameter and neglecting small terms as above, we obtain the following dispersion equation:

$$D(\omega,\mathbf{k}) = (q_{n-2y}\alpha_{n-2}q_{n-1y}\alpha_{n-1}q_{n+1y}\alpha_{n+1}q_{n+2y}\alpha_{n+2} - q_{n-1y}\alpha_{n-1}\beta_n^-\beta_{n+1}^+$$
$$-q_{n-2y}\alpha_{n-2}q_{n-1y}\alpha_{n-1}\beta_{n+1}^-\beta_{n+2}^+ - \beta_{n-2}^-\beta_{n-1}^+q_{n+1y}\alpha_{n+1}q_{n+2y}\alpha_{n+2})q_{ny}\alpha_n \quad (31)$$
$$-q_{n-2y}\alpha_{n-2}\beta_{n-1}^-\beta_n^+q_{n+1y}\alpha_{n+1}q_{n+2y}\alpha_{n+2} - q_{n-2y}\alpha_{n-2}\beta_n^-\beta_{n+1}^+q_{n-1y}\alpha_{n-1}q_{n+2y}\alpha_{n+2} = 0$$

Considering the solution of Eq. (31) near the synchronism point (19), we can rewrite the equation for $\delta\omega$ as

$$\delta\omega^2 \pm 2\Omega_{bn}\delta\omega = \frac{i}{2}h\frac{\Delta_b\omega_b^2\omega_n^2}{u^2\gamma_0^5}\frac{A}{B}e^{-2\frac{\omega_n b}{u\gamma_0}} = 0, \quad (32)$$

where

$$A = q_{n-2y}q_{n+2y}\left\{\left(1+\frac{\chi u\gamma_0^2}{\omega_n}\right)q_{n-1y}\beta_{n+1}^+ - \left(1-\frac{\chi u\gamma_0^2}{\omega_n}\right)q_{n+1y}\beta_{n-1}^+\right\} \quad (33)$$

and

$$A = q_{n-2y}q_{n-1y}q_{n+1y}q_{n+2y} - q_{n-2y}q_{n-1y}\beta_{n+1}^-\beta_{n+2}^+ - \beta_{n-2}^-\beta_{n-1}^+q_{n+1y}q_{n+2y}. \quad (34)$$

For $q_{n-1y} = 0$ the shift of frequency $\delta\omega$ does not depend on the small parameter $h$. This example shows that high orders of the surface modes can play a significant role for the SP instability and, therefore, they should be taken into account.

### 3. Dispersion equation: General form

Now, when we know the form of the dispersion equation for the SP instability, we can consider the general case. We assume that the surface of grating is described by an arbitrary periodic function $y = y(x)$ wrere $y(x) = y\left(x+\frac{2\chi}{n}\right)$ and $n$ is an integer. The metal grating is assumed to have ideal conductivity. This approach excludes the cases of rectangular grating and other gratings in which the form of the surface cannot be described by a function. But the results of this approach can be used for the case of an arbitrary profile of the grating. Below we will show how this approach can be used for the case of a rectangular grating. Substituting the solutions of Maxwell equations in the general boundary condition for metal surface

$$E_x(x, y(x)) + y'(x)E_y(x, y(x)) = 0, \qquad (35)$$

multiplying by $\exp(-ip\chi x)$, where $p = 0; \pm 1; \pm 2; \pm 3,...$ and then integrating over coordinate $x$ within the interval $(0, d)$ we obtain an infinite system of algebraic equations for the partial amplitudes $P_n$:

$$\sum_{n=-\infty}^{n=\infty} T_{np} P_n = 0, \qquad (36)$$

with the following coefficients

$$T_{np} = \left( q_{ny} + (n-p)\chi \frac{q_{nx}}{q_{ny}} \right) \left( G_{np}^+ [1 + K(n)] - G_{np}^- K(n) e^{2iq_{ny}b} \right). \qquad (37)$$

Here

$$G_{np}^\pm = \int_0^{2\pi} e^{i(n-p)\xi \pm q_{ny} y(\xi)} d\xi, \qquad (38)$$

and $\xi = \chi x$ is a dimensionless coordinate.

For a non-zero solution of Eqs. (36) to exist, the determinant of the system must be zero:

$$D(\omega, \mathbf{k}) = det \|T_{np}\| = 0. \qquad (39)$$

This is the most general form of the dispersion equation for the Smith-Purcell instability without resonator. Its roots give the spectrum of frequency and the growth rates. As above, the zero-order approximation for dispersion equation, giving the SP frequency spectrum, corresponds to the case with the plane mirror boundary of $y(x) = 0$, when $G_{np}^\pm(h=0) = 2\pi \delta_{np}$, and the dispersion equation takes the following form:

$$D_0(\omega, \mathbf{k}) = \prod_n q_{ny} \prod_n \alpha_n = 0. \qquad (40)$$

The value $\alpha_n$ has been early defined in (6). The solutions of $\alpha_n = 0$ give the spectrum of resonant frequencies (12). Without loss of generality, we assume that the mode of SP-spectrum

with n excites. Then, expanding the determinant on the sum over column n, Eq. (39) can be rewritten as:

$$D(\omega,\mathbf{k}) = \sum_{p=-\infty}^{p=\infty} T_{np} \hat{M}^{np} = 0, \qquad (41)$$

where $\hat{M}^{np}$ is the main minor, corresponding to the matrix coefficient $T_{np}$:

$$T_{np}(\omega-\omega_n)^2 = \left(q_{ny} + (n-p)\chi \frac{q_{nx}}{q_{ny}}\right) G_{np}^+ \left((\omega-\omega_n)^2 - \Omega_{bn}^2\right)$$
$$-\frac{\Delta_b \omega_b^2 \omega_n}{2u\gamma_0^4} e^{-2\frac{\omega_n b}{u\gamma_0}} \left(q_{ny} + (n-p)\chi \frac{q_{nx}}{q_{ny}}\right)(G_{np}^+ - G_{np}^-). \qquad (42)$$

Introducing the value

$$Z_n = \frac{\sum_{p=-\infty}^{p=\infty} \left(q_{ny} + (n-p)\chi \frac{q_{nx}}{q_{ny}}\right)(G_{np}^+ - G_{np}^-)\hat{M}^{np}}{\sum_{p=-\infty}^{p=\infty} \left(q_{ny} + (n-p)\chi \frac{q_{nx}}{q_{ny}}\right) G_{np}^+ \hat{M}^{np}}, \qquad (43)$$

we rewrite Eq. (41) as:

$$D(\omega,\mathbf{k}) = (\omega-\omega_n)^2 - \Omega_{bn}^2 - Z_n \frac{\Delta_b \omega_b^2 \omega_n}{2u\gamma_0^4} e^{-2\frac{\omega_n b}{u\gamma_0}} = 0. \qquad (44)$$

For frequency $\omega$ given by Eq. (19) we find $(\omega-\omega_n)^2 - \Omega_{bn}^2 = \delta\omega^2 \pm 2\Omega_{bn}\delta\omega$.

Introducing the following notation

$$\mu = \frac{Z_n}{e^{-2\frac{\omega_n b}{u\gamma_0}} - 1}, \qquad (45)$$

we obtain the final form of the dispersion equation for $\delta\omega$:

$$D(\omega,\mathbf{k}) = \delta\omega^2 \pm 2\Omega_{bn}\delta\omega + \mu_n \Omega_{bn}^2 = 0. \qquad (46)$$

The upper plus sign corresponds to the branch $\omega = \omega_n + \Omega_{bn}$, while the lower minus sign corresponds to the branch $\omega = \omega_n - \Omega_{bn}$. The solutions of Eq. (46) are $\delta\omega = \pm\Omega_{bn}\left(1 \pm \sqrt{1-\mu_n}\right)$. The complex frequency of SP instability is

$$\omega = \omega_n \pm \Omega_{bn}\sqrt{1-\mu_n}, \quad n=1,2,3,... \tag{47}$$

The growth rate of instability is $\delta\omega'' = \Omega_{bn}\,\text{Im}\sqrt{1-\mu_n}$. If $|\mu_n| \ll 1$, we obtain the collective regime of generation, when $|\delta\omega| \gg \Omega_{bn}$, with $\delta\omega \approx \pm\frac{i}{2}\Omega_{bn}\mu_n$. The growth rate is $\delta\omega'' = \pm\frac{1}{2}\Omega_{bn}\,\text{Im}(\mu_n)$.

For $|\mu_n| \gg 1$, we obtain the Thompson regime of generation, when $|\delta\omega| \square\, \Omega_{bn}$, with $\delta\omega \approx \pm i\Omega_{bn}\sqrt{\mu_n}$. The growth rate for the single-particle type is $\delta\omega'' = \pm\Omega_{bn}\,\text{Re}\left(\sqrt{\mu_n}\right)$.

If one branch of coupled frequencies, for example, the branch with high frequency, increases, then the other branch decreases and vice versa.

### 4. Conclusion

We have used the framework of the dispersion equation to study coherent Smith-Purcell (SP) radiation induced by a relativistic magnetized electron beam in the absence of a resonator. We have found that the dispersion equation describing the induced SP instability is a quadratic equation for frequency; and the zero-order approximation for solution of the equation, which gives the SP spectrum of frequency, corresponds to the mirror boundary case, when the electron beam propagates above plane metal surface (mirror). It was found that the conditions for both the Thompson and the Raman regimes of excitation do not depend on beam current and depend on the height of the beam above the grating surface. The growth rate of the instability in both cases is proportional to the square root of the electron beam current. No feedback is needed to provide the coherent emission.

We have presented a theoretical treatment of the small signal gain regime of a SP FEL. Considering an electron beam of infinite thickness being at a finite height above the grating, we have established the dispersion relation for the electromagnetic waves excited. An arbitrary grating is described with metal boundary condition. The dispersion relation has been studied considering amplification of the surface as well as radiation modes. An analytical expression for the growth rate of SP instability has been derived.